\documentstyle[12pt]{article}
\newcommand{\beq}{\begin{equation}}
\newcommand{\eeq}{\end{equation}}
\newcommand{\beqa}{\begin{eqnarray}}
\newcommand{\eeqa}{\end{eqnarray}}
\newcommand{\beqar}{\begin{eqnarray*}}
\newcommand{\eeqar}{\end{eqnarray*}}

\def \prob {{\cal P}}
\def \la {\langle}
\def \ra {\rangle}

\def \h {{\cal H}}

\begin{document}
\begin{titlepage}
\vspace{.5in}
\thispagestyle{empty}
\begin{flushright}
UBC-TP-95010
\end{flushright}
\vspace{0.1in}
\begin{center}
{\bf\Large First Order Corrections to the Unruh Effect}\\
\vspace{.3in}
B. Reznik
\footnote{\it e-mail: reznik@physics.ubc.ca}\\
\medskip
{  Department of Physics}\\
{ University of British Columbia}\\
{6224 Agricultural Rd. Vancouver, B.C., Canada
V6T1Z1}\\
\end{center}
\vspace{.2in}
\begin{center}
\begin{minipage}{5in}
\begin{center}
{\large\bf Abstract}

\end{center}

{\small
First order corrections to the Unruh effect are calculated
from a model of an  accelerated particle detector of  finite mass.
We show that quantum smearing of the trajectory and
large recoil essentially do not modify the Unruh effect.
Nevertheless,  we find corrections to the thermal distribution
and to the Unruh temperature.
In a certain limit, when the distribution at equilibrium
remains exactly thermal, the corrected temperature is found
to be $T = T_U( 1 - T_U/M)$, where $T_U$ is the Unruh temperature.
We estimate the consequent corrections to the Hawking
temperature and the black hole entropy, and comment on the relationship
to the problem of trans-planckian frequencies.}
\end{minipage}
\end{center}
\end{titlepage}

\section{Introduction}

In many respects  the Unruh effect
and the Hawking effect are
manifestations of the same phenomenon [1-3].
In both cases the accelerated detector (or asymptotic observer)
observes a thermal spectrum of particles.
The thermal radiation is associated with  information (or entropy)
which is hidden beyond an event horizon.
In both cases the thermal radiation  originates due to
an exponentially increasing red shift
between the rest frame
(or asymptotic infinity) and the initial Minkowski vacuum (ingoing
fields).
In fact, the
similarity between the two effect is much more than  formal. This can be
seen by examining the relationship   of the two effects near a black hole.
The existence of a Hawking flux of emitted radiation at infinity
depends on a suitable boundary condition at the horizon.
However a sufficient condition that  Hawking radiation will
be observed at infinity is that
a stationary particle detector located at a constant
$r$ near the horizon will detect a
thermal bath of radiation with the Unruh temperature
$T_U= {a\over 2\pi}$, where $a$ is the detector's proper acceleration.
The converse is also true; if Hawking radiation is seen at
infinity, this  implies that the stationary detector must see
the Unruh temperature near the horizon.
Therefore, to some extent, the red shifted
Hawking radiation
near the horizon  is a ``hot'' Unruh radiation.

The close relationship between the two effects suggests
that a better   understanding of  the role of quantum
effects in the case of an accelerated Unruh detector might
shed light on the case of the black-hole.
When the mass of the detector is taken to be finite, one can no longer
ignore the quantum mechanical smearing of the trajectory and
the recoil back-reaction when the detector is excited.
In the case of the black hole, the first effect of   quantum smearing
might be analogous to the  quantum
smearing  of the black-hole horizon, i.e. of the causal structure.
The second effect, that of the detector's recoil,
might be related to the back-reaction of the black hole
when a  Hawking quanta is emitted.
In this paper we shall study these effects for the case of the
Unruh detector and attempt to estimate the implications
for the case of the black hole.

In his original paper \cite{unruh}, Unruh suggested a two-field model
for a finite mass accelerated detector.
Two scalar fields $\Phi_M$ and $\chi_{M'}$,
of masses $M$ and $M'=M+\Omega$, respectively, were  taken to represent
two states of a detector of mass $M$ with two internal energy levels
with an energy gap $\Omega$. By introducing a coupling of the form
$\epsilon\phi\chi_{M'}\Phi_M$ with
a field $\phi$, this detector can detect  quanta associated with
the field $\phi$.
Recently, Parentani studied a similar two-field detector model
\cite{parentani}.
Using the WKB approximation
to describe the fields and a stationary phase approximation to
calculate the  transition amplitudes,  Parentani showed that when
the quantum smearing is smaller than the typical length $1/a$,
the Unruh effect is unmodified.

In this work,  the problem is approached by using another model.
In  Section 2 we present a model for a first quantized relativistic
particle detector that is accelerated  by a constant external electric field.
The geometry of the detector's trajectory is described
by introducing  future and past Rindler horizon operators \cite{qframe}.
We then compute in Section 3 the first order transition amplitude.
What we find is that a large quantum smearing in detector's trajectory
and the (possibly) exponentially large recoil of the detector
do not modify the Unruh effect.
Nevertheless, the recoil back-reaction does induce corrections
in the probability distribution at equilibrium
and in the Unruh temperature.
The origin  of these corrections is that different energy
levels of the detector experience different acceleration and hence
``see'' different temperatures.
We calculate the first order correction to the thermal spectrum.
Only in two  limits -- that  of $\Omega/T_U<<1$ and   $\Omega/T_U>>1$,
where $\Omega$ is the excitation energy --
does the probability
distribution remain  exactly thermal.
In the first limit, we once again obtain back the
Unruh temperature. However in the second limit, of $\Omega>> T_U$,
we find a correction to the Unruh temperature
 given by $T= T_U(1-T_U/m)$,
where $m$ is the detector's rest mass.

In Section 4 we study more qualitatively the nature
of the final state of the detector +
field system. Using the geometrical event horizon operators,
the final state is represented as an entangled state of field and
detector or horizon  states.
The back reaction can then be expressed as a shift in the
location of the Rindler  horizons.
The location of the horizons with respect
 to the initial state of the detector is shifted with respect to
the location of the horizons with respect to the final state.
This  shift can be exponentially large.

In the final section we attempt to apply our results to the case of
the black hole. Since the acceleration in this case
is determined by the black hole's mass (and not by the detector's mass
as is in the case of the electric field), the correction is
a genuine property of the black hole.
The connection between the Unruh temperature near the horizon,
and the Hawking temperature at infinity,
is used to extrapolated from the Unruh temperature
to a corrected Hawking temperature.
The modification of the latter leads to
a logarithmic correction to the black hole
entropy. We also comment on the relation of our results to the problem
of trans-planckian frequencies.

In the following we adopt the units in which $\hbar=k_B=c=G=1$.

\section{Accelerated Detector with  Finite Mass}

In this section we  present a model for a particle
detector of finite mass which takes into account
also the quantum nature of the detector's trajectory.

Consider a particle detector of rest mass $m_0$ and charge $q$
in a constant external electric field $E_x$ in $1+1$ dimensions.
Let us describe the internal structure by a harmonic oscillator
with a coordinate  $\eta$ and frequency $\Omega$.
The internal oscillator
is coupled to a  free
scalar field $\phi$.
The total effective action is
\beq
S = -m_0\int d\tau  -  qE_x\int X dt +
 {1\over2}\int\biggl(\Bigl({d\eta\over d\tau}\Bigr)^2
-\Omega^2\eta^2\biggr) d\tau + \int g_0 \eta \phi(X(t(\tau)),t(\tau))d\tau
+ S_F.
\eeq
Here, $\tau$ is the proper time in the detector's rest frame,
$X$ is the position of the detector,
$g_0$ is the coupling strength with
a scalar field $\phi$ and $S_F$ is the action of the field.
Since we would like to describe the back reaction on the trajectory
let us rewrite this action in terms of the
 inertial frame time $t$. The action of the
accelerated detector is then given by
\beq
 \int \biggl[\Bigl(-m_0 - g_0\eta\phi(X,t)  \Bigr)\sqrt{1-\dot X^2} -
qE_x X\bigl] dt    +  {1\over2}\int\biggl[{1\over\sqrt{1-\dot X^2} }
\Bigl({d\eta\over dt}\Bigr)^2 - \sqrt{1-\dot X^2}
\Omega^2\eta^2 \biggr]dt.
\eeq
This yields a simple expression for the Hamiltonian of the total
system with respect to the inertial frame:
\beq
H = \sqrt{P^2 + M^2}  -  qE_xX + H_F,
\label{1}
\eeq
where the effective mass $M$ is given by
\beq
M= m_0 + {1\over 2} \Bigl( \pi_\eta^2 + \Omega \eta^2 \Bigr)
+ g_0 \eta \phi(X),
\eeq
and $\pi_\eta = {\partial L \over \partial\dot \eta} = \dot \eta/
 \sqrt{1-\dot X^2}$.
The validity of our model rest upon a the assumption that the Schwinger
pair creation effect can be neglected for our detector.
Since the the Schwinger pair creation process
is damped by the factor  $\exp(-\pi M^2/qE_x)$ this
implies the limitation $M^2>qE_x$. Notice that since
the acceleration is $a=qE_x/M$, this implies that
$M> a = 2\pi T_U$.
In the following we set $E_x=1$ for convenience.

To obtain a quantum mechanical model we simply need to
impose quantization conditions  on the
conjugate pairs $X, P$ and $q, \pi_q$ and use the standard
quantization procedure for the scalar field.
It is convenient to introduce internal energy level
raising and lowering operators $A^\dagger$ and $A$.
The harmonic oscillator Hamiltonian
can then be replaced by $\Omega A^\dagger A\equiv \Omega N$
and the internal coordinate
by $\eta=i(A^\dagger-A)/\sqrt{2\Omega}$.
This form can also be used in other, more general cases, however the
simple commutation relation  $[A, A^\dagger]=1$
in the case of a harmonic oscillator, needs to be modified
accordingly.

So far we have not imposed a limitation on the
coupling strength $g_0$. In the
case of small coupling $g_0(t) = \epsilon(t)$ the
Hamiltonian can be written to first order in $\epsilon(t)$ as
\beq
H = H_D - qX  + H_F + H_I.
\eeq
Here
\beq
H_D = H_D(P, N) = \sqrt{P^2 + (m_0+\Omega A^\dagger A) ^2}
\eeq
is the free detector Hamiltonian,
$H_F$ is the free field Hamiltonian
\beq
H_F= {1\over2}\int dx' [\Pi_{\phi}^2 + (\nabla\phi)^2+m_f^2\phi^2 ],
\eeq
and
\beq
H_{I} =
 i\epsilon(t)  \biggl\lbrace  {m_N\over H_D},
(A^\dagger-A) \phi(X,t) \biggr\rbrace,
\eeq
where $m_N \equiv m_0 +N\Omega= m_0 +\Omega A^\dagger A$ and
 the anti-commutator,
$\lbrace A,B\rbrace ={1\over2}(AB+BA)$, maintains hermiticity.
We have also absorbed a factor of $1/\sqrt{2\Omega}$ in the definition
of $\epsilon(t)$.
Comparing this interaction term with
that used in the absence of a back-reaction we note that
apart from the  appearance of an  anti-commutator there
is also a new factor ${m_N\over H_D}$. As we shall see,
it corresponds to an operator boost factor  from the
inertial rest frame to the detector's rest frame.

In the Hiesenberg representation   the eqs. of motion for the
detector's coordinates $X$ and $P$ are given by:
\beq
\dot X = {P\over H_D}
 - i\epsilon(t)  \Bigl\lbrace {m_N P\over H_D^3},
(A^\dagger-A)\phi(X) \Bigr\rbrace,
\label{X}
\eeq
\beq
\dot P = q
-i\epsilon(t)  \biggl\lbrace  {m_N \over H_D},
(A^\dagger-A) \phi'(X,t) \biggr\rbrace,
\label{P}
\eeq
where $\phi' = {\partial\phi\over\partial x}$.
We also have
\beq
\dot A = -i  \bigl(H_{D,N+1} - H_{D,N}\Bigr)A  -i[A, H_I],
\label{A}
\eeq
and
\beq
(\Box - m_f^2)\phi(x,t) =
 i\epsilon(t) \biggl\lbrace  {m_N\over H_D},
 (A^\dagger-A) \delta(x-X) \biggr\rbrace .
\eeq

In the zeroth order approximation ($\epsilon=0$)
the solution of eqs. (\ref{X}-\ref{A}) is
\beq
X^{(0)}(t)= X_0 +{1\over q} \Bigl[H_D(t)- H_D(t_0)\Bigr], \ \ \ \
P^{(0)}(t) = P_0 +q(t-t_0),
\eeq
\beq
H_D^{(0)}(t) = \sqrt{ (P_0+q(t-t_0))^2 +(m_0+\Omega  A^\dagger_0 A_0)^2},
\eeq
and
\beq
A^{(0)}(t) =  \exp \Bigl[
-i\int_{t_0}^t ( H_{D,N_0+1}- H_{D,N_0})dt'  \Bigr]A_0.
\label{At}
\eeq
Here the subscript was used to denote the operator at
time $t=t_0$ and the superscript to denote
the zeroth order solution. To simplify notation we shall drop
the superscript.
Notice that $N_0 = A_0^\dagger A_0$ is a constant of motion in the zeroth
order approximation.

It is now useful to introduce a proper time $operator$ $\tau(t)$:
\beq
\tau = \int_{t_0}^t {m_0+\Omega A^\dagger A \over H_D} dt =
{m_0+\Omega A^\dagger A \over q}
\sinh^{-1}\Bigl[{q(t-t_0)+ P_0\over m_0+\Omega A^\dagger A}\Bigr].
\label{taut}
\eeq
We see that the factor  $({m_0+\Omega A^\dagger A)/H_D}= {m_N/ H_D}$
appearing in eq. (\ref{taut}) is the operator boost factor
${d\hat \tau(t)\over dt}$, from the inertial frame to the
detector's rest frame. Notice that $\tau$ depends only on $P_0$ and $N$.

In terms of the proper time operator, the detector's trajectory can
be simplified to:
\beq
t- t_0 - \tilde T_0 = {1\over a}\sinh a\tau,
\label{tt}
\eeq
\beq
X-\tilde X_0 = {1\over a} \cosh a\tau,
\label{xt}
\eeq
where
\beq
\tilde T_0 = -{P_0 \over q} \ \ \ \ \ \ \ \tilde X_0 = -{H_D\over q},
\eeq
and  the acceleration $a$ is given by the $operator$
\beq
a = a_N= {q\over m_0 + \Omega A^\dagger A} = {q\over m_N}.
\eeq
The operators $\tilde T_0$ and $\tilde X_0$  determine
the location  of the Rindler coordinate system of the detector with
respect to
the Minkowski coordinates ($t,x$). The space-time
location of the intersection point
of the future and past Rindler horizons is given by
$(-t_0 -\tilde T_0, -\tilde X_0)$. Since
\beq
[\tilde X_0, \tilde T_0 ] = {i\hbar\over q},
\label{TXtilde}
\eeq
the location of this  space-time point becomes quantum mechanically
smeared.

Another set of useful operators \cite{qframe} we shall introduce is that
of the location
of the future and past Rindler horizons $\h_+$ and $\h_-$, respectively.
They can be found from the relations
\beq
\h_+(t)=\lim_{t\to \infty}X(t), \ \ \ \ \ \ \ \
\h_-(t) =\lim_{t\to-\infty}X(t).
\eeq
We find
\beq
\h_+(t) = -\tilde T_0 +\tilde X_0 +t-t_0 = {P(t)\over q} - {H_D\over q},
\label{hp}
\eeq
and
\beq
\h_-(t) = \tilde T_0 +\tilde X_0 -(t-t_0) = -{P(t)\over q}- {H_D\over q},
\label{hm}
\eeq
Therefore we can express $X(t)$ as
\beq
X(t) = \h_+(t) + {1\over a} e^{-a\tau} \ \
 \stackrel{{t\to\infty}}{\to} \ \ \h_+(t),
\eeq
and
\beq
X(t) = \h_-(t) +  {1\over a} e^{a\tau} \ \
\stackrel{{t\to-\infty}}{\to} \ \ \h_-(t).
\eeq
In terms of $\h_\pm$, the Hamiltonian of the detector in
an external electric field has the simple form:
\beq
H_{acc} = H_D - qX = -{q\over 2} \Bigl( \h_+ + \h_- \Bigr).
\eeq
Finally, $\h_\pm$ satisfy the commutation  relation:
\beq
[\h_-, \h_+] = 2{i\hbar\over q}
\label{hpm}
\eeq
Examining eqs. (\ref{TXtilde}) and (\ref{hpm}), we
notice that since $q =  a m$,
in the limit of  constant
acceleration but large mass, the commutators
vanish as $m^{-1}$ and the classical
trajectory limit is restored.

\section{The Transition Amplitude}

We shall now proceed to calculate the first order
transition amplitude between the internal energy levels  $n$
and  $n+1$ of the detector.
To this end it will be most convenient to  use
the interaction representation.
The operators in this representation are the solutions of the
free equations of motion given by (\ref{At},\ref{taut},\ref{tt},\ref{xt}),
and the wave function satisfies   the Schr\"odinger equation
\beq
i\partial |\Psi\ra = H_I |\Psi\ra.
\eeq
Given at $t=t_0$ by the initial wave function
$|\Psi_0\ra$, to first order in $\epsilon$
the final state at time $t$ is given by
\beq
|\Psi(t)\ra = \biggl[1-i\int_{t_0}^t \
\epsilon(t')  \biggl\lbrace  {m_0+\Omega A^{\dagger}A\over H_D},
i(A^{\dagger}-A) \phi(X,t') \biggr\rbrace dt'  \biggr]
|\Psi(t_0)\ra.
\eeq

Let us set initial conditions for the internal oscillator
to be in the $n$'th exited
state $|n\ra$,
and for the scalar field to be in a
Minkowski vacuum state $|0_M\ra $.
The initial
state of the total system is therefore given by
$|\Psi(t_0)\ra = |0_M\ra\otimes|n\ra\otimes|\psi_D\ra$,
where $|\psi_D\ra$
denotes the space component of the  detector's wave function.
Using the solution (\ref{At}) for $A$ and $A^\dagger$,
the transition amplitude can be expressed as:
\beq
|\Psi(t)\ra = |\Psi(t_0)\ra  -{\epsilon\over2}\int_{t_0}^t dt'\biggl[
\eeq
$$
 \sqrt{n+1}|n+1\ra
\biggl( {m_{n+1}\over H_{D,n+1}}
e^{i\int_{t_0}^{t'}\Delta H_{n+1}dt''}
\phi(X_n,t')+
e^{i\int_{t_0}^{t'}\Delta H_{n+1}dt''}
\phi(X_n,t'){m_n\over H_{D,n}} \biggr)
$$
$$
-\sqrt{n}|n-1\ra
\biggl( {m_{n-1}\over  H_{D,n-1}}
e^{-i\int_{t_0}^{t'}\Delta H_{n}dt''}\phi(X_n,t')+
e^{-i\int_{t_0}^{t'}\Delta H_{n}dt''}\phi(X_n,t'){m_n
\over H_{D,n}} \biggr)
$$
$$
\biggr]\otimes|0_M\ra\otimes|\psi_D\ra.
$$
Here we used the notation
$\Delta H_{n} = H_{D,n}- H_{D,n-1}$.
 The subscript $n$ (e.g. in $X_n$),  means that we need to substitute
the free solutions with  $N=n$.
In two dimensions the solutions for a free massless scalar field
can always be separated into right and left moving waves, i.e.
$\phi = \phi_L(V)+ \phi_R(U)$ where $U=t-x, \ \ V=t+X$.
For simplicity we will limit the discussion to massless
scalar fields and examine
the solution  only for right moving waves.
Therefore, we substitute for $\phi$:
\beq
\phi_R(U) = \int {d\omega\over \sqrt{4\pi\omega}} \Bigl(e^{-i\omega U}
a_\omega + e^{i\omega U}a^\dagger_\omega \Bigr).
\eeq
Using eqs. (\ref{tt},\ref{xt},\ref{hp}) we find that on the trajectory of the
detector the light cone coordinate $U$ is given by
\beq
U|_{D} = t- X = -\h_{+0} -{1\over a} e^{-a\tau}.
\eeq
The final state can be written as:
\beq
|\Psi(t)\ra = |\Psi(t_0)\ra
-{\epsilon\over2}\int{d\omega\over\sqrt{4\pi\omega}}
 \int_{t_0}^t dt'\biggl[
\label{tamp}
\eeq
$$\sqrt{n+1}|n+1\ra
\biggl( {m_{n+1}\over H_{D,n+1}}e^{i\int\Delta H_{n+1}dt''}
e^{i\omega(-\h_{+0n} -{1\over a_n} e^{-a_n\tau_n})}
+
e^{i\int\Delta H_{n+1}dt''}
e^{i\omega(-\h_{+0n} -{1\over a_n} e^{-a_n\tau_n})}
{m_n\over H_{D,n}} \biggr)
$$
$$
-\sqrt{n}|n-1\ra
\biggl( {m_{n-1}\over H_{D,n-1}}e^{-i\int\Delta H_{n}dt''}
e^{i\omega(-\h_{+0n} -{1\over a_n} e^{-a_n\tau_n})}
+
e^{-i\int\Delta H_{n}dt''}
e^{i\omega(-\h_{+0n} -{1\over a_n} e^{-a_n\tau_n})}
{m_n \over H_{D,n}} \biggr)
$$
$$
\biggr]\otimes|1_{\omega M}\ra\otimes|\psi_D\ra.
$$

This is an exact result in the first order approximation
in $\epsilon$. So far we have not introduced
additional assumptions on $m_0$, $\Omega$ or $a_n=q/m_n$.
We shall now apply a large mass limit.  We shall assume that
\beq
m_0  >> a_0 = {q\over m_0}.
\eeq
This restriction is indeed equivalent
to a suppression of the
Schwinger pair production process.
Since the Unruh radiation has temperature
$T_U=a/2\pi$ we shall need only energy gaps with $\Omega\sim a$.
therefore we can also set
\beq
m_0 > n\Omega,
\label{small-omega}
\eeq
where $n= O(1)$.
Under these assumptions we can simplify the terms in (\ref{tamp}).
First consider the term $\exp(\int \Delta H_{n+1}dt)$.
Using (\ref{small-omega}) we expand:
\beq
i\int_{t_0}^t \Delta H_{n+1}dt' =
i\Omega\int_{t_0}^t {m_n\over H_{D,n}}
\Bigl[
1+{1\over2} {\Omega\over m_n}{P^2\over H_{D,n}^2}\Bigr] + O(\Omega^3/m^3)
\eeq
$$
= {i\Omega\tau_n} + {i\over2}\Omega^2
\Bigl[{1\over m_n}\tau_n - {1\over q}\tanh(a_n\tau_n) \Bigr]
+ c(P_0)+ O(\Omega^3/m^3)
$$
$$
\simeq
 i\Omega\tau_n\Bigl(1+ {1\over2} {\Omega\over m_n}
- {a_n\over q}\exp(-a_n\tau_n)\Bigr) +c(P_0)+ O(\Omega^3/m^3),
$$
where $c(P_0)$ is a constant, and  in the last line  we have used the large
$\tau$ approximation.
This approximation is justified since the transition amplitude is
dominated by contributions arising from integration over
large $\tau$.
In the following we shall hence neglect the exponential correction
and the constant $c(P_0)$ which gives rise only to an overall
phase, and use the approximation:
 \beq
i\int_{t_0}^t \Delta H_{n+1}dt' =
i\Omega\tau_n\Bigl(1+ {1\over2} {\Omega\over m_n} \Bigr).
\label{intdh}
\eeq

Next consider the exponential terms in (\ref{tamp})
which contain the horizon operator  $\h_+$.
Only these terms maintain a dependence on
the operator $X$ as  $\h_{+0} = X + G(P_0)$, where $G$ is a function of $P_0$.
Using the Baker-Hansdorff identity we obtain:
\beq
\exp[{-i\omega(\h_{+0} +{1\over a} e^{-a\tau})}]=
\exp\Bigl[
{{i\over2 q}\omega^2 e^{-a\tau}{m_n\over H_{D,n}} +O({1\over q^2})}
\Bigr]
\exp({-i\omega{1\over a} e^{-a\tau}})
\exp({-i\omega\h_{+0}})
\label{exph}
\eeq
The $O(q^{-2})$ corrections will be neglected in the following.
Notice that since $[\h_+, P_0] = i\hbar/q$, the unitary
operator $e^{-i\omega\h_{+0}}$ generates  the translation:
 $p_0 \to p_0 +\omega$. In other words, this unitary operator
generates the recoil which is required to conserve the total
momentum when the detector is exited and
a scalar Minkowski photon is emitted.

Finally, we consider the boost operator:
\beq
{m_{n+1}\over H_{D,n+1}} = {m_{n}\over H_{D,n}}
\biggl[1 +{\Omega\over m_{n}}
\Bigl(1-{m_n^2\over H_{D,n}^2}\Bigr) + O(\Omega^2/m^2)
\biggr].
\eeq
Since for large $\tau$
\beq
{m_n\over H_{D,n}}= {1\over \cosh(a_n\tau_n)} =
 2 e^{-a\tau_n} - O( 2e^{-3a\tau_n}),
\eeq
we  shall approximate this boost factor by
\beq
{m_{n+1}\over H_{D,n+1}} =  {m_{n}\over H_{D,n}}
\biggl[1 +{\Omega\over m_{n}} \biggr].
\label{bf}
\eeq

We can now return to the transition amplitude (\ref{tamp})
and for simplicity focus only on the  amplitude  $A(\omega,n+1,p)=
\la 1_\omega, n+1, p_0|\Psi(t)\ra$
using eqs.  (\ref{intdh},\ref{exph},\ref{bf}) we find
\beq
A(\omega,n+1,p_0)=
 -{i\epsilon\over2}\sqrt{n+1\over 4\pi\omega}
\int^t_{t_0} dt'
\biggl[
\biggl({m_{n}\over H_{D,n} (p_0+\omega)}
+{m_{n}\over H_{D,n}(p_0)}\Bigl(1+{\Omega\over m_n}\Bigr)\biggr)\times
\label{int}
\eeq
$$
\exp\bigg(i\Omega\Bigl(1+{1\over2}{\Omega\over m_n}\Bigr)
\tau_n -         {i\omega{1\over a} e^{-a\tau_n}}
+{{i\over2 q}\omega^2 e^{-a\tau_n}{m_n\over H_{D,n}} }
\biggr)\biggr]\phi_D(p_0+\omega).
$$
Here, $\phi_D(p)= \la p|\psi_D\ra$.
To obtain (\ref{int}) we  used a representation with  $\h_{+0}$ and $P_0$
as conjugate operators, and acted with the unitary operator
$\exp-i\omega\h_{+0}$ to generate translations in the momentum.
At this point the
transition amplitude is expressed as a c-number integral.

Let us proceed to investigate this integral.
For large $t$
the phase $\theta$ of the integrand can be approximated by
\beq
\theta= \Omega\Bigl(1+{1\over2}{\Omega\over m_n}\Bigr)
\tau_n -  \omega{1\over a} e^{-a_n\tau_n}
+{1\over q}\omega^2 e^{-2a_n\tau_n}.
\label{phase}
\eeq
The stationary phase condition yield
\beq
\omega \simeq - \Omega\Bigl(1- {\Omega\over 2m_n} \Bigr)
 e^{a_n\tau_n}.
\label{sphase}
\eeq
This can be compared with the case of a classical trajectory
obtained by sending $m\to \infty$.
In the present case, the  frequency at the stationary point
is shifted. However, with  the assumption
${\Omega\over m_n}<1$, the correction is small and this
 frequency  remains exponentially high.

Next notice that the recoil affects only one of the boost factor
 ${m_{n}\over H_{D,n}(p_0+\omega)}$ in eq.  (\ref{int}),
by a shift of the momentum.
This has a simple physical interpretation.
The transition amplitude is a superposition of two terms which correspond
to two different ``histories''. In one history, the detector is first
boosted by $\omega$ and only then it ``absorbs'' a scalar photon.
In the second term, the detector first absorbs a photon and
only afterwards it is boosted. Therefore in this term the boost factor
is not affected by the recoil.

The shift of  $p_0\to p_0 +\omega$ in the boost factor,
is equivalent to a shift in time given by
$t\to t' = t+{\omega\over e}$. In terms of the proper time (which is
now a c-number) this
correspond to the transformation
\beq
\tau \ \ \to \ \ \tau' = \tau + {\omega\over q}e^{-a\tau}
\eeq
For  transitions with $\tau(t)-\tau(t_0)>>1/a$,
this transformation does not
modify the integral. Hence in terms of $\tau'$:
\beq
{m_{n}\over H_{D,n}(p_0+\omega)}= {d\tau'_n\over dt}.
\eeq

The second, unshifted,  boost factor can be expressed in terms of $\tau'$ as
\beq
{m_n\over H_{D,n}}\Bigl(1+ {\Omega\over m_n}\Bigr)=
{d\tau'\over dt}\Bigl( 1
+{1\over m}(\Omega + \omega e^{-a\tau'}) +O(\Omega^2/m_n^2) \Bigr).
\eeq
Hence by expressing the
integral (\ref{int}) in terms of  $\tau'$ we find that the
two terms are equal up to order $O(\Omega^2/m^2)$ and an
additional piece  that (up to this order) vanishes
at the stationary point (\ref{sphase}).

Expressing the phase in terms of $\tau'$ we find
\beq
\theta=\Omega\Bigl(1+{1\over2}{\Omega\over m_n}\Bigr)\tau_n'
 -{\omega\over a_n} \Bigl(1+{\Omega\over m_n}\Bigr)
e^{-a_n\tau_n'} +O(\Omega^2/m^2).
\eeq
where the term involving ${\omega^2\over q}e^{-2a\tau}$
in eq. (\ref{phase}) has dropped out and we
are left only with the higher order corrections
$O(\Omega^2/m^2)$, which we will neglect.

In terms of $\tau'$ the amplitude $A(\omega,n+1,p_0)$ can be written as:
\beq
 -{i\epsilon}\sqrt{n+1\over 4\pi\omega}\phi_D(p_0+\omega)\Biggl[
\int d\tau_n'\exp\bigg({i\Omega\Bigl(1+{1\over2}{\Omega\over m_n}\Bigr)\tau_n'}
-i\omega{1\over a}\Bigl(1+{\Omega\over m_n}
\Bigr) e^{-a\tau_n'}  \biggr) + {\xi\over m_n} \Biggr]
\eeq
where
\beq
\xi ={1\over 2} \int d\tau(\Omega +\omega e^{-a\tau})\exp\Bigl(
i\Omega\tau - i{\omega\over a}(1+ {\Omega\over m})e^{-a\tau} \Bigr)
\eeq
For large $\tau $,  $\xi\sim O({\Omega\over m})$, and
the term $\xi/m$ can be
neglected.

Finally we obtain
\beq
A(\omega,n+1,p_0) = {i\epsilon}\sqrt{n+1\over 4\pi\omega}\phi_D(p_0+\omega)
a_n^{-1}
\Bigl(  {\omega\over a_n}(1+ {\Omega\over m_n})\Bigr)^{i{\Omega'\over a_n}}
\Gamma(-i{\Omega'\over a_n}) e^{-{\pi\Omega'\over 2a_n}}  +O(\Omega^2/m^2),
\eeq
where $\Gamma$ is the Gamma function, and
\beq
\Omega' = \Omega\Bigl( 1+ {1\over2} {\Omega\over m_n} \Bigr).
\eeq

Comparing this amplitude to that obtained in the case of
a fixed classical trajectory,
we notice that it appears to be modified  only by a
pure phase factor and by the shift $\Omega\ \ \to \ \ \Omega'$.

To first order, the transition probability is therefore given by
\beq
\prob(\omega,n\to n+1,p_0) =
 {\epsilon^2(n+1)\over 4\pi\omega}
|\phi_D(p_0+\omega)|^2 \biggl({2\pi\over a_n\Omega'}\biggr)
{e^{-{\pi\Omega'/a_n}}\over e^{\pi\Omega'/a_n} -  e^{-{\pi\Omega'/ a_n}} }
\label{tprob}
\eeq
This expression has the same form as of a
a thermal transition probability
 with a shifted Unruh temperature $T'_U$:
\beq
T_U'(n) = {q\over m_0 + n\Omega'}\simeq T_U\Bigl
(1 -{n\Omega^2\over m_n^2}\Bigr) .
\label{ttemp}
\eeq

Nevertheless,  this  transitions probability
does not imply a thermal  distribution at equilibrium.
Notice that the ``transition'' temperature  (\ref{ttemp})
depends on the energy level $n$.
Indeed, since the acceleration
depends on $n$, each energy state of the detector
 ``sees'' a slightly different temperature.
The temperature gradient between two neighboring levels is given by
$\Delta T_n/T_n = \Omega/m_n$.

In order to find the distribution at equilibrium we
 need to compare the probability of
excitation,  $\prob(\omega,n+1\to n, p_0)$, to
the transition probability for de-excitation,
 $\prob(\omega,n+1\to n, p_0)$.
By examining eq. (\ref{tamp}) we find that up to corrections
of $O(\Omega^2/m^2)$,
the  de-excitation probability $\prob(\omega, n+1\to n,p_o)$ is obtained
by the substitution  $a_n\to a_{n+1}$ and $\Omega'\to -\Omega'$
in eq. (\ref{tprob}).
Using the approximation $a_{n+1} = a_n(1-\Omega/m_n)$
we find
that the probability distribution at equilibrium satisfies,
up to a correction of
$O(\Omega^2/m^2)$, the relation
\beq
\prob(n\to n+1) = \Bigl(1+{\pi\Omega\Omega'\over  a_nm_n}
\coth(\pi\Omega'/a_n) \Bigr)
\exp\Bigl(
-{2\pi\Omega' \over a_n}(1+{\Omega\over2m_n} + {a_n\over2\pi m_n})
\Bigr)  \prob(n+1\to n),
\label{equi}
\eeq
which is stisfied for every $\omega$ and $p_0$.

This probability distribution can be simplified in two limiting cases.
For   $\Omega<<T_U$ we get back the ordinary  thermal relation
\beq
\prob(n\to n+1) = \exp\Bigl(-{\Omega\over T_U}
)\prob(n+1\to n),
\label{bolz}
\eeq
where $T_U= a_n/2\pi$. This should have been anticipated.
In this limit,  the temperature gradient $\Delta T/T$ between
 nearby energy levels, vanishes.

The more interesting limit is obtained for $\Omega> T_U$.
In this limit we obtain back an exact thermal distribution:
\beq
\prob(n\to n+1) = \exp\Bigl(-{\Omega\over T_{acc}}\Bigr)\prob(n+1\to n).
\label{bolz2}
\eeq
However the Unruh temperature receives a correction:
\beq
T_{acc} = {a_n\over 2\pi}\Bigl(1 - {a_n\over 2\pi m_n}\Bigr) =
T_U\Bigl( 1 -  {T_U\over m} \Bigr)
\label{ctem}
\eeq

By repeating the stages of this calculation it can be verified that the
same correction to the probability distribution and to the temperature
is also obtained from the transition amplitude involving left
moving photons, i.e. from the interaction with the part $\phi_L(V)$
of the scalar field. Therefore it seems that in this limit
eq. (\ref{ctem})  constitutes a genuine first order
correction to the Unruh temperature.
Since in higher levels the effective acceleration is smaller,
this correction indeed acts to reduce the Unruh temperature.

It is interesting to notice that even in the limit of
 ${\Omega\over m} << 1 $, when the  Unruh temperature is
restored, the recoil back-reaction can be still large.
The recoil shifts the momentum by $\omega$, which by eq. (\ref{sphase})
can be exponentially large even for small $\Omega$.
A further  discussion of this  recoil back-reaction
and of the  quantum smearing effect is given in the next section.


\section{Recoil and Quantum Smearing}

In this section we examine the back-reaction effect
on the trajectory of the detector.
Let us re-state the results of the last section in a more qualitative
way. For the case of a classical trajectory, it was shown by
Unruh and Wald \cite{uw} that if the detector is initially in  the
ground state then the
final state can be written as
\beq
|\Psi(t)\ra = |\Psi_0\ra -i|n=1\ra\otimes
a_{R\Omega} |0_M\ra.
\eeq
Here, $a_{R\Omega}$ is the annihilation
operator of a quantum with frequency $\Omega$
with respect to the Rindler coordinate
system that is defined by the detector's trajectory.
Using the well known relation \cite{unruh} of $a_{R\Omega}$
to Minkowski  creation and annihilation operators
$a_{M}$ and $ a_{M}^\dagger$,
they get
\beq
|\Psi(t)\ra = |\Psi(0)\ra -i C(\Omega,a)|n=1\ra
{e^{-\pi\Omega/a} \over
(e^{\pi\Omega/a} -e^{-\pi
\Omega/a})^{1/2}} a_{M}^\dagger |0_M\ra,
\label{uw}
\eeq
where  $C$ is a normalization factor.
Note that $a^\dagger_{M}$ creates a positive frequency  Minkowskian
photon,  which is not in a state of definite frequency $\omega$.
Qualitatively we can use the stationary phase approximation
eq. (\ref{phase}) to relate the typical frequency
of this photon to the time of emission $\tau$.

We can now  use the result obtained  in the last section
to replace eq. (\ref{uw}) with
\beq
|\Psi(t)\ra = |n=0,\psi_D,0_M\ra
\eeq
$$-i C(\Omega,a_n)|n=1\ra
{e^{-2\pi\Omega/a} \over
(e^{\pi\Omega/a} -e^{-\pi
\Omega/a})^{1/2}}\Bigl(
e^{-iH_F \h_{+}}a_{M R}^\dagger + e^{+iH_F \h_{-}}
a_{M L}^\dagger
\Bigr)
|0_M,\psi_D\ra
\label{shift}
$$
Here we ignored the effect of temperature gradient between different
energy levels, which gave rise to the correction found in the
previous section. We have also
restored the full coupling with the left and right moving
waves.
The operators $a_{M R}^\dagger $ and $a_{ML}^\dagger$,
 correspond to creation operators of  right and left moving waves
 respectively.
This equation can be easily generalized to the case of transitions between
any two levels $n$ to $n+1$, as well as to the case of de-excitations.
We have assumed that the scalar field is massless. However, for
a massive field
we simply need to replace $e^{-iH_f \h_{+0}}$ by $e^{iH_f\tilde T_0 -iP_f
\tilde X_0}$ etc.

The new feature of eq. (\ref{shift}) is
the insertion of the horizon shift operators $\exp( \pm i H_F \h_\pm)$
which act on the wave function of the detector and of the scalar field.
These shift operators generate correlations between
the ``emitted'' Minkowski scalar photon and the trajectory of the
detector.

To illustrate  these correlations, let us concentrate only on the
left moving  waves and express $a_{MR}^\dagger$ in terms of creation
operators of definite  Minkowski frequency:
\beq
a_{MR}^\dagger = \int f(\omega) a_\omega^\dagger d\omega
\eeq

Eq. (\ref{shift}) can now be written as
\beq
|\delta\Psi\ra = -iC'|n=1\ra\int d\omega dh_{+} e^{-i\omega h_{+}}
f(\omega)\psi(h_{+}) |1_\omega\ra\otimes|h_{+}\ra.
\label{ent}
\eeq
Here we used a basis of
$\h_{0+}$: $\h_+|h_+\ra = h_+|h_+\ra$.
We see that the recoil interaction generates
correlation between the shift
$h_+$ in the $u$-time of the right moving ``emitted''
Minkowski photons with the ``horizons states'' $|h_+\ra$ of
the detector. Therefore, the
effect of ``smearing the horizon'' yields after emission the
 final entangled state (\ref{ent}).
In each component of this state,
the Unruh effect is manifested, with the correction
discussed in the previous section.
Since the corrections  do not depend on the uncertainty
or the smearing $\Delta h_+$ of the future event horizon, the overall
wave function still manifests the Unruh effect.

In order to examine the effect of the emission on the detector we
can re-write eq. (\ref{ent}) by using as a basis the past horizon
operator $\h_-$. We obtain:
\beq
|\delta\Psi\ra= -iC'|n+1\ra\int d\omega dh_-
f(\omega)\psi(h_-) |1_\omega\ra\otimes|h_--\omega\ra,
\eeq
where $\psi(h_-) = \la h_-|\Psi_D\ra$.
Since $\h_\pm$ are conjugate operators, the  operator
$\exp-i\omega\h_+$  has shifted the past horizon operator by $\omega$.
It is interesting  to notice that the shift by $\omega$
of the past horizon can be exponentially large.
In fact, from the stationary phase approximation we get that
it is related to the time of emission $\tau$ as:
$\Omega\simeq \Omega\exp (a\tau)$.
Therefore a  detection of a particle of energy $\Omega$
generates an exponential shift in the location of the past horizon
of the detector:
\beq
\delta h_- = h_{-out} -h_{-in}  \simeq \Omega \exp (a\tau)
\label{dhm}
\eeq
The meaning of this shift is as follows.
We can use the initial state $\psi_{in}$ to define the location
$h_{-in}$ of the past horizon.
We can also use the final state $\psi_f$ of the detector
and by propagating it
to the past (with the free Hamiltonian)
determine the location $h_{-out}$.
These two locations differ by an exponential shift.

The use of propagation of a wave function to the past might seem
strange. However the same phenomenon occurs if the
detector is excited in the past at $\tau<0$.
In this case it emits a left moving Minkowski particle.
We find that this induces an
exponentially large shift in the location of the future event
horizon operator
$\h_+$:
\beq
\delta h_+ = h_{+out} - h_{-in} \simeq \Omega \exp (-a\tau)
\label{dhp}
\eeq

The manifestation of the back reaction as an exponentially large
shift is related to the method  of
't Hooft \cite{thooft} and of  Schoutens, Verlinde and   Verlinde
\cite{verlinde}. In their case,
infalling matter into the black hole,
induces an exponential shift of
the time of emission of the Hawking photon in the future.
The reason is that the Hawking photons stick so close
to the horizon that even a small shift of the horizon
still modifies the time of emission.
In our case this exponential shift is related to the
exponential energy of the emitted Minkowski photon.
In both cases, the back reaction requires the existence of exponentially
high frequencies in the vacuum.
As in the case of Hawking radiation, a naive cutoff
eliminates the thermal spectrum seen by the Unruh detector.

\section{Correction to Black Hole Radiation}

As noted in the introduction, the Unruh effect and the Hawking
effect are very closely related. It is therefore
conceivable that the same type of corrections are relevant
in both cases.
Let us recall how the Unruh effect is manifested in the
case of the black hole.
A  fiducial particle detector (i.e. stationary at  constant angles and
Schwarzschild
radius $r$),
in the gravitational field of a black hole
will in general observe radiation.
Only in two limiting cases, that of $r$ going  to infinity and
of $r$  close to  the horizon, does this radiation take a simple form.
In the first case, the detector observes at spatial infinity
Hawking radiation with temperature $T_H= 1/8\pi M$,
where $M$ is the mass of the black hole. In the second case,
the detector will see the  Unruh radiation with temperature
 \cite{unruh}
\beq
T_U(r) = {a(r)\over 2\pi} \simeq
{1\over 8\pi M} {1\over \sqrt{g_{00}(r)}}=
{T_H\over \sqrt{g_{00}(r)}},
\label{tubh}
\eeq
where $a(r)$ is the proper acceleration  at a constant radius $r$.
This equation relates the Unruh temperature seen
very near to the horizon and the Hawking temperature at
$r>>2M$.

The origin of the  Unruh radiation, i.e. the proportionality of
the temperature to the acceleration near the horizon,
can be seen as follows.
The Schwarzschild metric
\beq
ds^2 = \Bigl(1-{2M\over r}\Bigr) dt^2 -
 \Bigl(1-{2M\over r}\Bigr) ^{-1}dr^2 - r^2d\Omega^2
\eeq
can be approximated near the horizon
by the metric
\beq
ds^2 = \Bigl({\rho\over 4M}\Bigr)^2dt^2 - d\rho^2 - r^2(\rho)
d\Omega^2,
\eeq
where
\beq
\rho \equiv 4M\Bigl(1-{2M\over r}\Bigr)^{1/2}
\simeq \int_{2M}^r {dr\over \sqrt{1-{2M\over r}}}
\eeq
is approximately the proper distance of the point $r$ to the
horizon.
Therefore near the horizon, the  reduced
2-dimensional Schwarzschild metric with coordinates
$t-\rho$   can be approximated by a
Rindler metric.
%
World lines with  constant $\rho(r)$ correspond to trajectories of
constant proper acceleration $a= 1/\rho$.

Consider now a particle detector of mass $m$ which is held by
an external force at a fixed radius $r$.
What is the nature of the corrections to the Unruh effect
in this case?
First we notice that contrary to  the case studied in the previous
sections, the acceleration (\ref{tubh}) in this case
is a function of the black hole mass, and not of the
mass of the detector.
Since we have seen that the corrections originate from  modifications
to the acceleration, we can expect that  in the present case
they are determined by the back reaction on the black hole.

When an Unruh particle is detected, the rest frame mass
of the detector is modified from $m$ to
$m+\delta m$.
If the total mass at infinity is unchanged
by this process, the excitation must be accompanied by a
decrease of the back hole's  mass by
\beq
\delta M = - \delta m \sqrt{g_{00}(r_D)}.
\label{dm}
\eeq
In other words, an excitation of the detector corresponds
to an emission by the black hole. The emitted photon
is absorbed by the detector and therefore
the total black-hole
and detector mass, remains unchanged as seen from infinity.
Nevertheless,
due to the back reaction effect (\ref{dm}),
the proper distance $\rho$ between the horizon and the
detector, or the acceleration $1/\rho$ of the detector have been
changed.

We have seen in Section 3. that the
correction to the Unruh temperature arises from the decrease
of the acceleration due to an absorption.
The correction (\ref{ctem})
was in fact due to the ratio  $a_{n+1}/a_n=(1+\delta a_n/a_n)$.
In the present case the decrease by $\delta M$ in the mass of the
black hole causes to a decrease in the proper acceleration.
Using eqs. (\ref{tubh}) and (\ref{dm}) we obtain
\beq
\delta a(r)  = -{\delta M\over 4M \sqrt{g_{00}(r)}}
\biggl( {1\over M} - {1\over r g_{00}(r)} \biggr)
\eeq
$$
\simeq {\delta M\over 8M^2 g_{00}^{3/2}} = -
{\delta m\over 8M^2 g_{00}}.
$$

Therefore,
\beq
{a_{M+\delta M}\over a_{M}}
 \simeq 1- 4\pi \delta m  T_U(r) \simeq
\exp\Bigl(-4\pi\delta m T_U(r)\Bigr).
\eeq
The Boltzmann factor is hence shifted to
\beq
\exp -{\delta m\over T_U(r)}\Bigl(1+4\pi T_U^2(r)\Bigr).
\eeq
This corresponds to a correction of the Unruh temperature:
\beq
T_{acc}(r) = T_U(r)\Bigl( 1 - 4\pi T_U^2(r) \Bigr).
\label{taccbh}
\eeq
Indeed, since the back reaction effect is larger when the
 detector is closer to the horizon
the correction increases accordingly.
We note that the correction (\ref{taccbh}) tends to decrease the
temperature. This effect becomes large only as
the detector is lowered  to a distance of $\rho \sim l_{pl}$.
However, at this point we can no longer trust our
model since  in this limit the detector's mass needs to be of
order of $T_U(r)$.

What does eq. (\ref{taccbh}) imply for an observer at infinity?
When the back reaction is ignored,
the Unruh temperature  near the horizon,
is seen at infinity as red shifted
to the Hawking temperature:
\beq
\sqrt{g_{00}}T_U(r) = T_H.
\label{trans}
\eeq
In our case however, we can not simply multiply eq. (\ref{taccbh})
by the red shift factor in order to obtain the correction
to the Hawking
temperature.
The reason is that the source of this correction is  the modification
in $g_{00}(r_D)$ which is   cause by the back reaction effect.
Therefore,  by using the  classical metric we will obtain the
answer that the corrected Hawking temperature depends on the location of the
detector, which is of course incorrect.

Since we still expect only a small correction to the transformation
(\ref{trans}), it seems suggestive to extrapolate (\ref{taccbh})
by the substitution $T_U(r) \to  T_H$.
Therefore,
\beq
T_{BH}= T_H\Bigl( 1 - 4\pi T_H^2 \Bigr).
\label{ubh}
\eeq
The correction to  the Hawking temperature is very small for $M>>1$.
Although (\ref{ubh}) was obtained only for Hawking particles of
energy $E>T_H$,  it can still be a good estimate for the modification
for the complete spectrum. This allows us to obtain,
using the first law of thermodynamics,
the corresponding modification in the black hole entropy.
Up to an additive constant we obtain:
\beq
S_{BH} = 4\pi M^2 + {1\over 2} \ln M .
\label{entropy}
\eeq

We shall  conclude with a few remarks.
One of the main motivations for studying this problem of
a quantum Unruh detector was the hope that the quantum smearing
or the back reaction effects would render the problem
of exponentially high  trans-planckian frequency manageable \cite{jacobson}.
For small accelerations, it seems  that
nevertheless these new effects this problem is still unavoidable in our
model.
Only for a large accelerations or temperature of the order of
the detectors mass does the back reaction have
a significant effect. At this limit however,
the validity of our model breaks down.
It is interesting however, that if we naively extrapolate
the corrections to  the regime of large acceleration,
i.e., lower the detector nearer to the horizon,
the Unruh temperature decreases toward  zero instead
of diverging to infinity.
Such an effect could indicates an effective cutoff of
high frequencies near the horizon which
does not eliminate the Hawking effect.

\vspace {3cm}

{\bf Acknowledgment}

I wish to thank W. G. Unruh for  discussions and very helpful
comments. I have also benefited from discussions with
  S. Nussinov and J. Oppenheim.

\vfill \eject

\end{document}